# Experimental demonstration of enhanced slow and fast light by forced coherent population oscillations in a semiconductor optical amplifier

Perrine Berger,[1,2,*] Jérôme Bourderionnet,[1] Guilhem de Valicourt,[3] Romain Brenot,[3]
Fabien Bretenaker,[2] Daniel Dolfi,[1] and Mehdi Alouini[1,4]

[1]Thales Research & Technology, Campus Polytechnique, 1 avenue Augustin Fresnel, 91767 Palaiseau Cedex, France
[2]Laboratoire Aimé Cotton, CNRS-Université Paris Sud 11, Campus d'Orsay, 91405 Orsay Cedex, France
[3]Alcatel-Thales III-V laboratories, Campus Polytechnique, 1 avenue Augustin Fresnel, 91767 Palaiseau Cedex, France
[4]Institut de Physique de Rennes, UMR CNRS 6251, Campus de Beaulieu, 35042 Rennes Cedex, France
*Corresponding author: perrine.berger@thalesgroup.com

We experimentally demonstrate enhanced slow and fast light by forced coherent population oscillations in a semiconductor optical amplifier at gigahertz frequencies. This approach is shown to rely on the interference between two different contributions. This opens up the possibility of conceiving a controllable rf phase shifter based on this setup. © 2010 Optical Society of America

Slow and fast light in a semiconductor optical amplifier (SOA) enables realization of rf phase shifters, which are key components in several microwave photonics applications [1,2]. At gigahertz frequencies, the achievable phase shift experienced by an optically carried rf signal passing through the SOA has been recently increased up to $\pi$ by optically filtering out the redshifted modulation sideband before detection [3]. Moreover, a phase shift of $2\pi$ at 19 GHz has also been obtained with this method by cascading several SOAs [4]. However, this method involves the use of a very sharp optical filter and is, consequently, efficient mainly for relatively high frequencies (typically above 5 GHz). Furthermore, the insertion of the optical filter leads to a significant noise enhancement [5]. A new interesting way to enhance the phase shift has recently been proposed by Anton et al. [6]. These authors theoretically studied an SOA whose current is modulated. Seminal works about a modulated SOA have already been investigated, in particular in the context of optoelectronic mixing in SOAs [7]. However, Anton et al. predict for the first time that modulating the injected current of the SOA can enhance the phase shift at a given frequency. The principle of such forced coherent population oscillations has been experimentally validated in an erbium-doped fiber amplifier at very low frequency (20 Hz) [8]. In this Letter, we experimentally demonstrate these forced coherent oscillations in an SOA at gigahertz frequencies. Moreover, we propose a simple physical interpretation of this new way to enhance coherent population oscillation (CPO)-induced slow and fast light.

A typical CPO is produced by a modulated optical power $P_{opt} = P_0 + Me^{-i2\pi f t} + \text{c.c.}$ incident on the SOA. This creates a modulation of the carrier density at frequency $f$, which dephases the optically carried rf modulation. Here, we consider the so-called forced coherent population oscillations, in which the current $I$ is also modulated at the same rf frequency $f$: $I = I_0 + I_1 e^{-i2\pi f t + i\phi} + \text{c.c.}$, with a phase shift $\phi$ with respect to the modulation of $P_{opt}$ [6]. Both sources of modulation produce a modulation of the carriers and, thus, of the complex gain. Consequently, the rf gain $G = \frac{M(L)}{M(0)}$ of a small slice of length $L$ of the SOA can be expressed as $G = \langle G \rangle + G_f e^{-i2\pi f t} + \text{c.c.}$, with $\langle G \rangle$ as the average gain and $G_f$ as the complex modulation amplitude of the gain at frequency $f$. The average gain $\langle G \rangle$ is the real saturated optical gain, which does not depend on $f$, as shown by the dashed blue curves in Fig. 1(a). As soon as the modulation index $I_1/I_0$ of the current reaches, typically, 0.1%, the contribution to $G_f$ coming from saturation is negligible with respect to the one induced by the current modulation. Consequently, owing to the finite carrier lifetime $\tau_s$, $G_f$ behaves like a first-order low-pass filter, as evidenced by the red dotted curves in Fig. 1(a). The overall gain $G$, which is the sum of the blue and red curves of Fig. 1(a), is the result of the interference between these two terms, as shown in Fig. 1(b). Because the argument of $G_f$ varies from $\phi$ at low frequency to $\phi + \pi/2$ at high frequency, the result of these interferences can take several shapes around a transition frequency $f_t$ [see Fig. 1(b)]. The frequency $f_t$ of this transition between the two filter shapes depends on $I_1$, $I_0$, and $P_0$. More precisely, $f_t$ increases with $I_1$. Consequently, if the frequency $f_t$ is higher than $1/\tau_s$, the phase shift will be maximal and will reach 180° if the phase difference $\phi$ between the two modulating signals is 90°: indeed, the phase of $G_f$ will reach 180° at $f_t$ [Fig. 1(b)]. This configuration, which is strictly equivalent to the configuration with a constant current, but with filtering out the redshifted sideband [3], is now going to be observed.

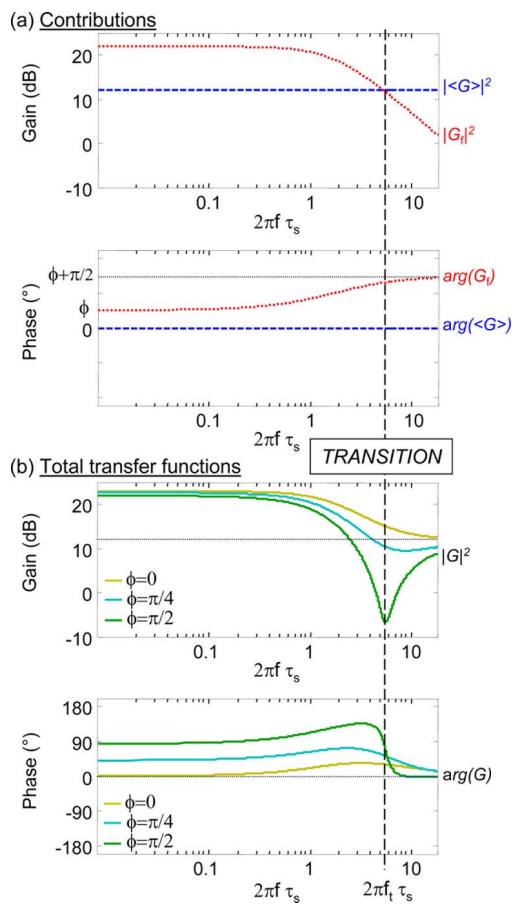

Fig. 1. (Color online) (a) Gain and phase of the major contributions. (b) Resulting rf gain and phase shift for different phase differences $\phi$ between the two modulating signals.

High-speed directly modulated SOAs are not commercially available. To experimentally demonstrate forced CPOs, we used a reflective SOA developed by the Alcatel-Thales III-V laboratory and specifically designed to be modulated at high frequencies [9]. The experimental setup is shown in Fig. 2(a). The microwave signal generated by the vector network analyzer (VNA) modulates both the injected current of the SOA and the input optical power. An rf attenuator enables us to control the modulation depth of the injected current. The phase difference between the two modulating signals is maintained at 90°. In the present experimental proof of concept, because of the delay between the two paths followed by the modulation, we select the frequencies for which the phase difference between the two modulations is equal to 90°. Of course, in a real implementation of this device, one would balance the paths and use a balanced hybrid 0°–90° divider instead of the power divider used here in conjunction with unbalanced arms. The current of the SOA is set to 80 mA. A calibration is done with the VNA when the SOA is disconnected. The gain and the phase shift introduced by the SOA are then measured for different modulation depths of the injected current, by introducing an attenuation $\eta$ on the modulation.

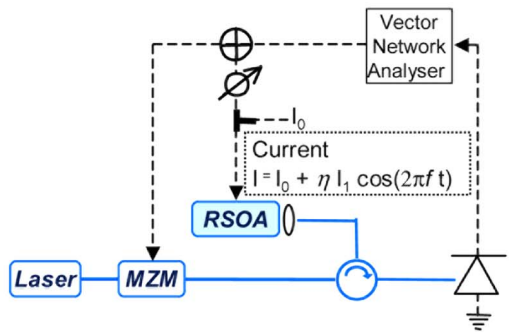
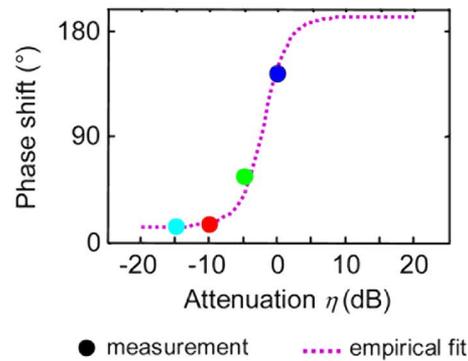
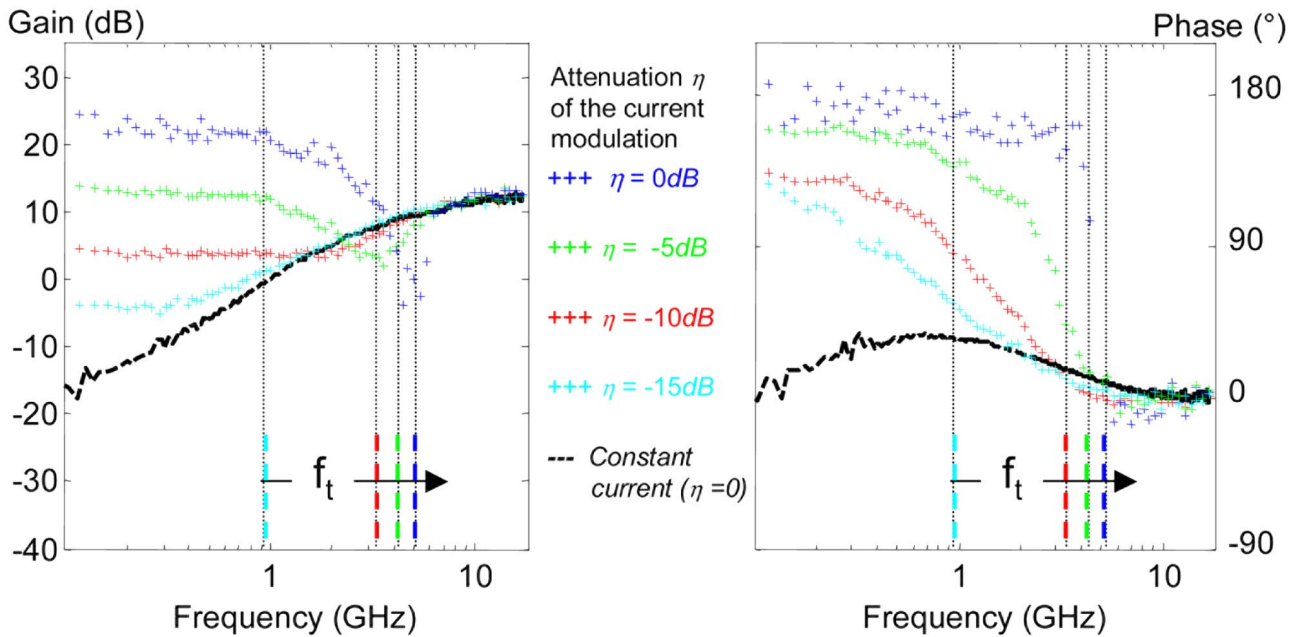

Fig. 2. (Color online) (a) Experimental setup: the rf signal generated by the VNA is divided by an rf power splitter. It is fed to the SOA current through a bias tee after a variable attenuator and is used to modulate the optical power through a Mach–Zehnder modulator (MZM). The photodetector restitutes the rf signal from the modulated optical signal that has traveled through the reflective SOA (RSOA) and the optical circulator. (b) Example of the use of our setup as a variable phase shifter. (c) Measurements of forced coherent population oscillations: rf gain (left) and phase shift (right) induced by the RSOA when the cw injection current is set to 80 mA. The different plots are obtained by varying the rf attenuation at the entrance of the SOA, that is, by varying the modulation depth of the injected current. The dashed black curve corresponds to the response of the SOA without any current modulation (standard CPO).

The results are shown in Fig. 2(c). Two different regimes can be identified with respect to the frequency. At high frequencies, the response (gain and phase) of the modulated SOA is similar to the response of the non-modulated SOA (usual CPO behavior, represented by the black dashed curve). In contrast, at low rf frequencies, forced coherent oscillations occur, and the phase tends to a value between 90° and 180°. This brings us to define a transition frequency, $f_t$, below which the phase of the signal can reach 180° and above which the phase is close to 0°. These observations confirm our physical interpretation. This transition is shown to be more or less sharp according to the modulation depth of the injected current. Moreover, it is seen that the $f_t$ can also be controlled by the modulation depth of the injected current; in Fig. 2(c), we tag the transition frequency $f_t$ for each attenuation. This behavior can be easily exploited to design a adjustable phase shifter, as shown in Fig. 2(b). In this example, the phase is controllable from 15° to 144° for a fixed frequency $f = 3.2$ GHz. These performances are similar to those achieved using sideband optical filtering before detection [3]. In our case, a $\pi$ phase shift could be achieved by increasing the current modulation depth. This could also increase the maximum operation frequency of the shifter. Indeed our simulations show that a larger ratio between the modulation depth of the current and the optical modulation index would lead to a phase shifter whose operating frequency is higher than 10 GHz.

To conclude, we have experimentally demonstrated for the first time, to the best of our knowledge, an enhancement of slow and fast light by forcing CPO in an SOA at gigahertz frequencies. We have shown that it can be explained using a simple physical interpretation. A transition between two regimes has been pointed out: at low frequencies, forced CPO due to the modulated current dominates, while, at high frequencies, optical gain overcomes. We show that the transition between these two phenomena is frequency tunable. This new physical interpretation highlights the underlying phenomena and convincingly explains the experimental observations. We show that forced CPO can be used to design a phase shifter when the interference between these two phenomena is destructive. These results are very similar to the observations done when the red-shifted modulation sideband is filtered out, with equivalent available phase shift, but without noise enhancement due to the filter (AM/FM conversion), nor wavelength dependence. A comparison of the underlying concepts of these two ways of enhancing slow and fast light will be pursued in a following paper. Finally, the experimental results presented here will be used as a basis in order to include forced CPO in the predictive models of gain, phase, noise, and nonlinearities we have recently proposed [10,11].

The authors acknowledge partial support from the DGA/MRIS, the GOSPEL EC/FET project and the ICT-FUTON project. The authors thank Francisco Arrieta Yáñez for fruitful discussions.